\documentclass{phbauth}        
\usepackage{graphicx}

\newcommand{\beq}{\begin{equation}}
\newcommand{\eeq}{\end{equation}}
\newcommand{\beqa}{\begin{eqnarray}}
\newcommand{\eeqa}{\end{eqnarray}}

\newcommand{\ii}{{\rm i}} 
\newcommand{\dd}{{\rm d}} 
\newcommand{\simg}{\stackrel{>}{_\sim}}
\newcommand{\siml}{\stackrel{<}{_\sim}}

\begin{document}

\begin{frontmatter}

\vspace{-25mm}

\title{
Phonon softening and double-well potential formation due to electron-phonon interaction in heavy-fermion systems
}

\author{Keisuke Mitsumoto\thanksref{thank1}} and 
\author{Yoshiaki \=Ono\thanksref{thank2}}

\address{Department of Physics, Niigata University, Ikarashi, Niigata 950-2181, Japan}

\vspace{-5mm}

\thanks[thank1]{E-mail: mitsumoto@phys.sc.niigata-u.ac.jp}
\thanks[thank2]{E-mail: y.ono@phys.sc.niigata-u.ac.jp}

\begin{abstract}
We investigate the periodic Anderson-Holstein model by using the dynamical mean-field theory combined with the exact diagonalization method. For the strong electron-phonon coupling $g\simg g_c$, the system shows an anomalous heavy-fermion behaviour which is accompanied by a large lattice fluctuation and an extreme phonon softening. We also calculate an effective potential for the ions and find that a simple harmonic potential for $g\siml g_c$ changes into a double-well potential for $g\simg g_c$. The effective pairing interaction between the conduction electrons shows a maximum at $g \approx g_c$ where the superconducting transition temperature is expected to be maximum. 
\end{abstract}

%

\end{frontmatter}



The A15 compounds such as $\mathrm{V_{3}Si}$ and $\mathrm{Nb_{3}Ge}$ have long been attracted much interest as they show high $T_{\rm c}$ and high $H_{\rm c2}$ superconductivity as well as anomalously large resistivity and Debye-Waller factor. 
Yu and Anderson\cite{Anderson1} originally proposed a local electron-phonon model where the strong electron-phonon coupling causes an effective double-well potential for the ion which is responsible for the anomalous behaviour observed in A15 compounds. As a strong coupling fixed point, the two-level Kondo systems were investigated to describe a heavy-fermion like behaviour observed in such compounds\cite{MatsuMi1}. More recently, the local electron-phonon model have been extensively studied by using the NRG approach\cite{KusuMi,Yotsuhashi}, but periodic (lattice) models were not discussed there.

Recently, another interesting class of materials has been observed in the filled skutterdites such as $\mathrm{PrOs_{4}Sb_{12}}$\cite{Goto} and the clathrates such as $\mathrm{Ce_{3}Pa_{20}Ge_{6}}$\cite{Nemoto}, where the rare-earth ion shows a rattling motion under a potential with several off-center minima. 
With the new findings, theoretical studies on a periodic Anderson model coupled with local phonons, {\it i.e.}, the periodic Anderson-Holstein model are highly desirable. 
The purpose of this paper is to present the results of the dynamical mean-filed theory (DMFT) for the periodic Anderson-Holstein model to elucidate the effect of the strong electron-phonon coupling on the heavy-fermion behaviour, the effective potential for the ions and the superconductivity.


Our Hamiltonian is given by
\begin{eqnarray}
\label{model}
H &=& \sum_{ij\sigma}t_{ij}c_{i\sigma}^{\dagger} c_{j\sigma}
       + \epsilon_{f}\sum_{i\sigma}f_{i\sigma}^{\dagger} f_{i\sigma}  \nonumber \\
  &+& V\sum_{i\sigma}(f_{i\sigma}^{\dagger} c_{i\sigma}+ h.c.)
      + U\sum_{i}n_{fi\uparrow} n_{fi\downarrow}  \nonumber \\
  &+& \omega_{0}\sum_{i}b_{i}^{\dagger} b_{i}
       + g\sum_{i}(b_{i}^{\dagger}+ b_{i})(\sum_{\sigma}n_{fi\sigma}-1),
\end{eqnarray}
where $c_{i\sigma}^{\dagger}$, $f_{i\sigma}^{\dagger}$ and
$b_{i}^{\dagger}$ are creation operators for a conduction ($c$)-electron  with spin $\sigma$ at site $i$, for a $f$-electron and for a phonon, respectively, and  $n_{fi\sigma}=f_{i\sigma}^{\dagger}f_{i\sigma}$. 
The quantities, $\epsilon_{f}$, $V$, $U$ and $g$, are the atomic $f$-level, the mixing between the $c$- and $f$-electrons, the on-site Coulomb interaction and the electron-phonon coupling strength. 
The density of $f$-electrons couples with the Einstein phonons whose frequency is $\omega_{0}$. 



To solve this model eq.(\ref{model}), we use the DMFT in conjunction with the exact diagonalization (ED) method\cite{Georges}. In the DMFT, the model eq.(\ref{model}) is mapped onto an effective single impurity Anderson-Holstein model\cite{Meyer,Koller}. 
Then, the local Green's function for a $f$-electron, 
\begin{equation}
G_{f\sigma}(\ii\omega_{n}) = -\int_{0}^{\beta}\langle T_{\tau}f_{i\sigma}(\tau)f_{i\sigma}^{\dagger}(0)\rangle e^{\ii \omega_{n}\tau}d\tau, 
\end{equation}
can be given by the impurity Green's function for the impurity Anderson-Holstein model. 
In the limit of infinite dimensions, the self-energy $\Sigma(\ii\omega_{n})$ becomes purely site-diagonal. 
Then the local Green's function for the $f$-electron satisfies  the following self-consistency condition:
\begin{eqnarray}
\label{self1}
G_{f\sigma}(\ii\omega_{n}) &=& \int d\epsilon\frac{\rho(\epsilon)}{\ii\omega_{n}-\epsilon_{f}-\Sigma(\ii\omega_{n})-\frac{V^{2}}{\ii\omega_{n}-\epsilon}} 
    \nonumber \\
   &=& \lbrack\tilde{G}_{f\sigma}(\ii\omega_{n})^{-1}-\Sigma(\ii\omega_{n})\rbrack^{-1},
\label{self2}
\end{eqnarray}
where $\rho(\epsilon)$ is the density of states (DOS) for the bare conduction band, $\tilde{G}_{f}(\ii\omega_{n})$ is the Green's function for the impurity Anderson-Holstein model with $U=g=0$ in an effective medium which will be determined self-consistently.
In addition, we obtain the local Green's function for the phonon,
\begin{equation}
D(\ii\omega_{n}) = -\int_{0}^{\beta}\langle T_{\tau}b_{i}(\tau)b_{i}^{\dagger}(0)\rangle e^{\ii\omega_{n}\tau}d\tau.
\end{equation}

To solve the impurity Anderson-Holstein model, we use the ED method for a finite-size cluster\cite{Koller,Caffarel}. 
All calculations are performed at $T=0$, and we replace the Matsubara
frequencies by a fine grid of imaginary frequencies $\omega_{n}=(2n+1)\pi/\tilde{\beta}$ with a fictitious inverse temperature $\tilde{\beta}$ which determines the energy resolution.

The numerical results for 8-site and $\tilde{\beta}=4000$ are almost the same as those for 6-site and $\tilde{\beta}=200$\cite{Ono}, therefore, we choose the parameters as 8-site and $\tilde{\beta}=4000$.
Also, the numerical results for 30 phonon states are almost the same as those for 100 phonon states\cite{Meyer}, therefore, we define the cutoff of phonon number is 30. 
In the following numerical results, we assume a semielliptic DOS for the bare conduction band with the bandwidth $W=1$, 
$\rho(\epsilon)=\frac{2}{\pi}\sqrt{1-\epsilon^2}$, and we set $\omega_0=0.05$ and $V=0.2$. 
We concentrate our attention on the particle-hole symmetric case with $\epsilon_{f}=-\frac{U}{2}$ where $\langle n_{c}\rangle=\langle n_{f}\rangle=1$.


\begin{figure}[h]
\begin{center}\leavevmode
\includegraphics[width=1.0\linewidth]{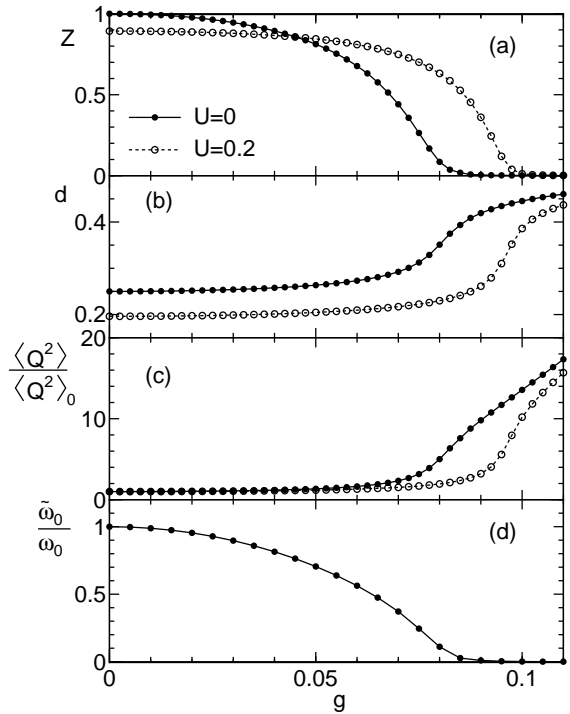}
\caption{ 
The quasiparticle weight $Z$, the double-occupancy $d$, the lattice fluctuation $\langle Q^{2}\rangle$  and the lowest excited energy of the phonon spectral function $\tilde{\omega}_0$ as functions of the electron-phonon coupling $g$ for $U=0$ (filled circles) and $U=0.2$ (open circles). 
$\langle Q^{2}\rangle_{0}$ and $\omega_{0}$ are the corresponding values for $g=0$. 
}\label{Fig1}\end{center}\end{figure}

In Fig.\ref{Fig1}(a), the quasiparticle weight, 
$Z=(1-\frac{\dd \Sigma(\omega)}{\dd \omega}|_{_{\omega=0}})^{-1}$,
is plotted as a function of $g$ for $U=0$ and $0.2$.
In both cases, $Z$ monotonically decreases with increasing $g$ and becomes extremely small but finite for the strong coupling $g\simg g_c$ where the mass enhancement factor $m^{*}/m=Z^{-1}$ becomes more than one hundred; $g_c\sim 0.085$ for $U=0$ and $g_c\sim 0.1$ for $U=0.2$ as shown in Fig.\ref{Fig1}(a). 
Thus we can conclude that the periodic Anderson-Holstein model eq.(\ref{model}) shows heavy-fermion behaviour due to the strong electron-phonon coupling in the realistic and wide parameter range. 
This is a striking contrast to the Holstein-Hubbard model in which the transition from the metal with $Z\ne 0$ ($g<g_c$) to the bipolaronic insulator with $Z=0$ ($g>g_c$) takes place at a critical value of the electron-phonon coupling $g_c$ and the heavy-fermion behaviour is observed only in the narrow range just below $g_c$\cite{Meyer,Koller}. 

In Fig.\ref{Fig1}(b), we plot the double-occupancy, $d=\langle n_{f\uparrow}n_{f\downarrow}\rangle$, as a function of $g$ for $U=0$ and $0.2$. 
When $g$ increases, $d$ gradually increases for $g\siml g_c$ while it dose steeply for $g\simg g_c$. 
The enhancement of $d$ is accompanied by the enhancement of the local charge fluctuation which causes the heavy fermion behaviour observed for $g\simg g_c$.

The lattice fluctuation is defined by 
$\langle Q^{2}\rangle=\langle \hat{Q}_i^{2}-\langle \hat{Q}_i\rangle^{2}\rangle$ 
with the lattice displacement operator, 
$\hat{Q}_i=\frac{1}{\sqrt{2\omega_{0}}} (b_i+b_i^{\dagger})$. 
Fig.\ref{Fig1}(c) shows the normalized lattice fluctuation, 
$\langle Q^{2}\rangle/\langle Q^{2}\rangle_{0}$, 
where $\langle Q^{2}\rangle_{0}=\frac{1}{2\omega_0}$ is the lattice fluctuation for $g=0$, {\it i.e.}, the zero-point oscillation. 
In the heavy fermion regime $g\simg g_c$, we can see the extreme enhancement of the lattice fluctuation which is accompanied by the enhancement of the local charge fluctuation together with the double occupancy shown in Fig.\ref{Fig1}(b). 


\begin{figure}[h]
\begin{center}\leavevmode
\includegraphics[width=0.9\linewidth]{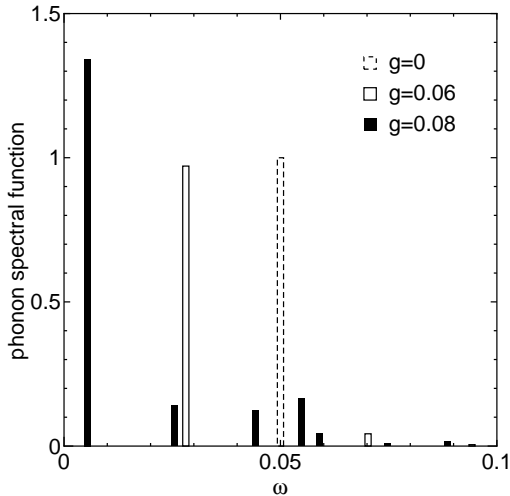}
\caption{ 
The phonon spectral function $-\frac{1}{\pi}\mathtt{Im}\mathnormal D(\omega+\ii 0_+)$ for several values of $g$ at $U=0$.
}\label{Fig2}\end{center}\end{figure}

In Fig.\ref{Fig2}, we plot the phonon spectral function, 
$-(1/\pi)\mathrm{Im}D(\omega+\ii 0_+)$, 
for $g=0$, $0.06$ and $0.08$ at $U=0$. 
We find that, with increasing $g$, the multi-phonon state appears and the lowest excited energy shifts to low energy. 
To see the energy shift in more detail, we show the lowest excited energy $\tilde{\omega}_0$ in the phonon spectral function as a function of $g$ in Fig.\ref{Fig1}(d). 
As shown in Fig.\ref{Fig1}(d), $\tilde{\omega}_0 =\omega_0$ for $g=0$ and $\tilde{\omega}_0$ decreases with increasing $g$ in proportion to $Z$ (see also Fig.\ref{Fig1}(a)). 
A remarkable soft phonon mode with $\tilde{\omega}_0 \approx 0$ is observed in the heavy fermion regime with $g\simg g_c$.

The large lattice fluctuation and the extreme phonon softening observed in Figs.\ref{Fig1}(c) and (d) can be explained by thinking in terms of an effective potential for the ions. 
To obtain the effective potential explicitly, we introduce a variational wave function for the ions\cite{Yotsuhashi},
\begin{eqnarray}
|\Psi_{v}(Q)|^{2} = A\exp[-(Bq^{2}+Cq^{4}+Dq^{6}+Eq^{8})] \label{vwf} 
\end{eqnarray}
with $q\equiv Q/Q_{0}$, where $Q$ is the lattice displacement and  $Q_{0}=\frac{1}{\sqrt{2\omega_0}}$ is that in the non-interacting case ($g=0$). 
In eq.(\ref{vwf}), the coefficients $A$, $B$, $C$, $D$ and $E$ are the variational parameters which will be determined to make 
$\langle Q^{2n}\rangle = \int Q^{2n}|\Psi_{v}(Q)|^{2} \dd Q $ ($n=0,1,2,3,4$) as close to $\langle Q^{2n}\rangle$ from the DMFT as possible. 
Then we define the effective potential for the ions as
\begin{equation}
V_{\rm eff}(Q)=\mathrm{log}[|\Psi_{v}(0)|^{2}/|\Psi_{v}(Q)|^{2}].
\end{equation}


\begin{figure}[h]
\begin{center}\leavevmode
\includegraphics[width=0.9\linewidth]{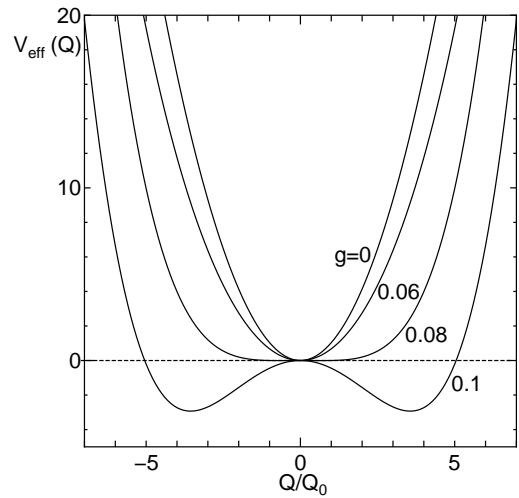}
\caption{ 
The effective potential for the ions $V_{\rm eff}(Q)$ for several values of $g$ at $U=0$.
}\label{Fig3}\end{center}\end{figure}

In Fig.\ref{Fig3}, $V_{\rm eff}(Q)$ is plotted for several values of $g$ at $U=0$. In the non-interacting case with $g=0$, the parameters are $A=\sqrt{\frac{\omega_{0}}{\pi}}$, $B=0.05$ and $C=D=E=0$ which actually yield the simple harmonic potential. 
With increasing $g$, $B$ monotonically decreases while $C$ increases to yield an anharmonic potential. In the heavy fermion regime with $g\simg g_c$, $B$ becomes negative while keeping $C$ positive to yield a double-well potential as shown in Fig.\ref{Fig3}. 
This behaviour has already been observed in the previous theories \cite{Anderson1,KusuMi,Yotsuhashi} where the local electron-phonon model was discussed in contrast to the present theory where the periodic (lattice) model is discussed. 
We note that the values of $D$ and $E$ are negligibly small and of orders of $10^{-8}\sim 10^{-10}$.

Finally, we discuss the superconductivity.
The renormalized local pairing susceptibility $\tilde{\chi}_{\rm loc}$ is given by\begin{eqnarray}
\label{chi}
\tilde{\chi}_{\rm loc} = -\int_{0}^{\beta}
\langle T_{\tau}c_{\downarrow}(\tau)c_{\uparrow}(\tau)
c_{\uparrow}^{\dagger}(0)c_{\downarrow}^{\dagger}(0)\rangle d\tau.
\end{eqnarray}
To calculate $\tilde{\chi}_{\rm loc}$ within the ED method, we use a spectral representation of r.h.s. in eq.(\ref{chi}) by inserting a complete set of eigenstates\cite{Ono}. We also calculate the bare local pairing susceptibility 
$\tilde{\chi}_{\rm loc}^{0}$ from the local Green's function $G_{\sigma}(\ii \nu)$ obtained from the DMFT,
\begin{equation}
\tilde{\chi}_{\rm loc}^{0} = -T\sum_{\nu}{G}_{\uparrow}(\ii\nu){G}_{\downarrow}(-\ii\nu). 
\end{equation}
By using $\tilde{\chi}_{\rm loc}^{0}$ and $\tilde{\chi}_{\rm loc}$, the local vertex function $\Gamma$ is given by
\begin{equation}
\Gamma=[\tilde{\chi}_{\rm loc}^{0}]^{-1}-[\tilde{\chi}_{\rm loc}]^{-1}. 
\label{Gamma}
\end{equation}
In the DMFT, the vertex function for the original lattice model can be given by the local vertex function for the effective impurity model\cite{Georges}.  Therefore, $\Gamma$ given in eq.(\ref{Gamma}) corresponds to the effective pairing interaction for the superconductivity.


\begin{figure}[h]
\begin{center}\leavevmode
\includegraphics[width=0.9\linewidth]{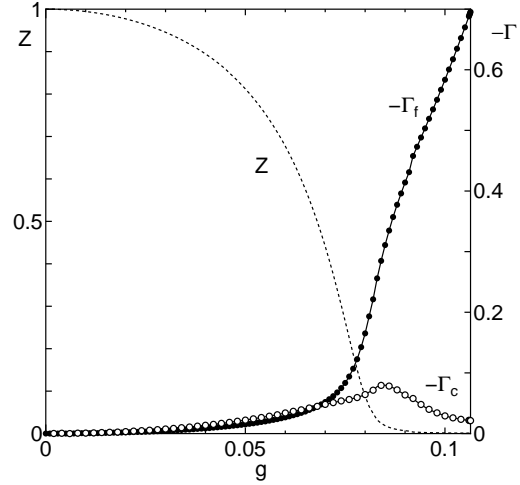}
\caption{ 
The vertex function for $f$-electrons $\Gamma_{f}$ (filled circles) and that 
for conduction electrons $\Gamma_{c}$ (open circles) 
together with the quasiparticle weight $Z$ (dotted line) 
as functions of $g$ for $U=0$. 
}\label{Fig5}\end{center}\end{figure}

In Fig.\ref{Fig5}, we plot the vertex function for $f$-electrons $\Gamma_{f}$ and that for conduction electrons $\Gamma_{c}$ together with the quasiparticle weight $Z$ as functions of $g$ for $U=0$. 
When $g$ increases, $-\Gamma_{f}$ increases in proportion to the lattice fluctuation $\langle Q^{2}\rangle$ (see also Fig.\ref{Fig1}(c)) and dose more rapidly for $g\simg g_c$. On the other hand, with increasing $g$, the renormalized band width for $f$-electrons decreases in proportion to $Z$ and becomes almost zero for $g\simg g_c$. 
From the both results of $-\Gamma_{f}$ and $Z$, we can suppose the superconducting transition temperature has a peak around $g_c$, as previously predicted by Matsuura and Miyake\cite{MatsuMi1} on the basis of the two-level Kondo lattice model. 
This is directly supported by the $g$-dependence of the pairing interaction between conduction electrons, $-\Gamma_{c}$, which has a maximum  at $g \approx g_c$ where the superconducting transition temperature is expected to be maximum.




The authors thank T. Goto, Y. Nemoto and K. Miyake for many useful comments and discussions. This work was partially supported by the Grant-in-Aid for Scientific Research from the Ministry of Education, Culture,  Sports, Science  and Technology.


\end{document}